# GPU-accelerated finite-temperature Lanczos method for spin Hamiltonians

Shadan Ghassemi Tabrizi[1,2*] and Thomas D. Kühne,[1,2]

[1]*Computational System Sciences, Technische Universität Dresden, 01187 Dresden, Germany*

[2]*Center for Advanced Systems Understanding (CASUS), Am Untermarkt 20, 02826 Görlitz, Germany, *s.ghassemi-tabrizi@hzdr.de*

**Abstract.** We present a GPU implementation of the finite-temperature Lanczos method (FTLM) for Heisenberg spin Hamiltonians that targets workstation hardware rather than distributed-memory clusters. The Hamiltonian action is evaluated matrix-free in a row-wise gather formulation. We introduce and compare two state-to-index strategies: a compressed lookup table (CLT), which reduces lookup memory by a factor of 16 relative to a full table while retaining a fixed, branch-light access pattern, and a GPU-adapted combinatorial-ranking scheme that removes the lookup table altogether. Numerical tests against FP64 CPU references show that FP32 GPU arithmetic changes heat capacities and magnetic susceptibilities by amounts several orders of magnitude below the stochastic uncertainty of the FTLM trace estimator at typical sample sizes. Benchmarks show speedups of up to about one order of magnitude over optimized multicore CPU calculations and enable Hilbert-space sectors of dimension $\sim 10^8$ on a single workstation GPU. The `MATLAB/CUDA` implementation, including example input files and benchmark scripts, is openly available at https://github.com/ghasdeke/ftlm-gpu (archived at DOI: 10.5281/zenodo.20378647) under the Apache-2.0 license.

## 1. Introduction

Finite-temperature properties such as magnetic susceptibility, heat capacity, and field-dependent magnetization are key experimental probes of exchange-coupled spin clusters and magnetic solids. Their quantitative modeling with spin Hamiltonians enables the extraction of effective exchange interactions from experimental data, and the exploration of thermodynamic behavior. For many relevant systems, including molecular magnets [1] and finite fragments of extended lattices [2], the Hilbert space is too large for full exact diagonalization (ED) but may remain accessible to iterative Krylov-space methods. Standard Lanczos calculations efficiently approximate extremal eigenvalues [3–5], whereas thermodynamic observables require traces over the full spectrum. The finite-temperature Lanczos method (FTLM) [6–8] addresses this problem by combining stochastic trace estimation over random states with Krylov-space approximations to thermal expectation values. Closely related typicality-based approaches obtain thermodynamic observables from a single thermal pure quantum state [9–11], and the kernel polynomial method provides an alternative spectral-density-based route to thermodynamic averages [12]. With suitable numbers of random vectors and Lanczos steps, FTLM yields accurate finite-temperature observables at a cost far below full diagonalization [13].

As explicit storage of the Hamiltonian in compressed sparse (CS) format becomes prohibitive for large spin clusters, matrix elements are commonly recomputed on the fly during sparse matrix-vector products (SpMV). The `spinpack` program [14,15] implements such direct Hamiltonian action following Ref. [16], enabling distributed-memory CPU calculations for state-space dimensions up to $\sim 10^{11}$ [17]. In this work, we show that workstation GPUs provide a complementary route. A matrix-free formulation is particularly well suited to GPU hardware, where high arithmetic throughput and massive thread-level parallelism can be used to compensate for limited memory capacity and bandwidth. Although the system sizes treated here do not rival the largest distributed-memory calculations, the proposed implementation makes FTLM calculations accessible on desktop hardware that would otherwise require dedicated CPU clusters.

GPU acceleration has been explored in quantum many-body physics and related fields. In ED, CUDA implementations of Lanczos-based solvers have produced substantial speedups for the Hubbard model [18]. More recently, scalable matrix-free SpMV implementations for distributed-memory ED have been demonstrated [19]. Related developments in quantum chemistry include full configuration interaction [20] and mean-field methods [21], with some implementations targeting consumer-grade hardware [22]. Established open-source frameworks for ED of correlated lattice models, such as HΦ [23], XDiag [24], and QuSpin [25], provide CPU-oriented reference implementations covering Lanczos, full diagonalization, and finite-temperature methods. The broader GPU-SpMV literature provides useful performance context [26,27], but FTLM for spin Hamiltonians has additional structure because the matrix elements are generated from local spin operations rather than read from a stored sparse matrix.

Here we present a MATLAB/CUDA implementation of FTLM for Heisenberg spin Hamiltonians on workstation GPUs, which evaluates the Hamiltonian action directly, without storing the sparse matrix, using the row-wise gather formulation of Ref. [16], and targets Hilbert-space sectors whose dimensions are limited primarily by GPU memory. Our main contributions are: (i) a compressed lookup table (CLT) for the state-to-index map that reduces memory by a factor of 16 relative to a full lookup table while retaining a fixed, branch-light access pattern suitable for GPU execution; (ii) a GPU-adapted combinatorial-ranking (CR) strategy that removes the lookup table when memory becomes limiting; (iii) a numerical precision analysis comparing FP32 GPU calculations with FP64 CPU references and showing that, for the thermodynamic observables considered here, FP32 deviations remain well below the stochastic FTLM uncertainty; and (iv) a CPU-GPU performance comparison for realistic FTLM workloads on representative antiferromagnetic Heisenberg clusters.

The paper is organized as follows. Section 2 summarizes the FTLM formalism and describes the matrix-free GPU kernel together with the CLT. Section 3 analyzes the numerical accuracy of FP32 arithmetic and benchmarks the GPU implementation. Section 4 summarizes the main results and discusses limitations and possible extensions.

## 2. Method

**Model.** We consider the Heisenberg model on a cluster of $N$ spin-$s$ sites, Eq. (1),

$$H = J\sum_{\langle i,j\rangle} \mathbf{s}_i \cdot \mathbf{s}_j = J\sum_{\langle i,j\rangle}\left[s_i^z s_j^z + \frac{1}{2}\left(s_i^+ s_j^- + s_i^- s_j^+\right)\right] \tag{1}$$

where the sum runs over the $N_\mathrm{B}$ bonds $\langle i,j\rangle$ of the cluster graph, and $J > 0$ corresponds to antiferromagnetic coupling. We work in the product basis $|m_1, m_2, \ldots, m_N\rangle$ with $m_i \in \{-s, -s+1, \ldots, s\}$. Since $[H, S^z] = 0$, where $S^z = \sum_j s_j^z$, the Hilbert space decomposes into sectors of definite total magnetization $M = \sum_i m_i$. Thermal averages thus reduce to sums of sector-wise traces (explicit computations can be restricted to $M \geq 0$, because sectors with $(+M)$ and $(-M)$ have identical spectra).

**FTLM.** These traces are estimated stochastically using a random-vector estimator [28,29]. For a Hilbert space of dimension $\mathcal{D}$, a set of $R$ normalized random vectors $|r\rangle$ gives Eq. (2) for the partition function (with $\beta = 1/k_B T$),

$$Z \approx \frac{\mathcal{D}}{R}\sum_{r=1}^{R}\langle r|e^{-\beta H}|r\rangle \tag{2}$$

and, for an observable $O$, Eq. (3),

$$O \approx \frac{\sum_{r=1}^{R} \langle r | O e^{-\beta H} | r \rangle}{\sum_{r=1}^{R} \langle r | e^{-\beta H} | r \rangle} \tag{3}$$

where the same set $\{|r\rangle\}$ is used in numerator and denominator. The stochastic uncertainty decreases with the number of random vectors as $1/\sqrt{R}$ and is controlled by the effective number of thermally occupied states (cf. [13] and references cited therein).

The matrix elements entering the traces are approximated in a Krylov subspace. Starting from $q_1 = |r\rangle$, Lanczos recursion without reorthogonalization yields the $N_L \times N_L$ symmetric tridiagonal matrix $T$. In exact arithmetic, $T = Q^T H Q$, where $Q = [\, q_1, \dots, q_{N_L} \,]$ is the Krylov basis. (In the implementation, this basis is not stored beyond the recursion; only the Lanczos coefficients defining $T$ are retained.) Diagonalizing $T$ gives the eigenpairs $(\theta_k, s_k)$. The first components $s_{k,1}$ of the normalized eigenvectors $s_k$ define the spectral weights, $w_k \equiv |s_{k,1}|^2$. In exact arithmetic, these weights equal the squared overlap between the starting vector and the corresponding Ritz vectors, $y_k = Q s_k$, i.e., $w_k = |\langle r | y_k \rangle|^2$. In finite precision, this identification is only approximate, because the Lanczos vectors need not remain orthogonal. However, the $\{s_k\}$ of $T$ are orthonormal, so the weights $|s_{k,1}|^2$ entering the FTLM estimator still satisfy the sum rule $\sum_k w_k = 1$.

For a single normalized random vector, the Lanczos approximation to the partition function is given by Eq. (4),

$$\langle r | e^{-\beta H} | r \rangle \approx \sum_k w_k e^{-\beta \theta_k} . \tag{4}$$

The corresponding Boltzmann-weighted energy moments are approximated in the same way: Eq. (5),

$$\langle r | H^p e^{-\beta H} | r \rangle \approx \sum_k w_k \theta_k^p e^{-\beta \theta_k}, \quad p = 1, 2 . \tag{5}$$

Combining the stochastic trace estimate with the Lanczos approximation yields Eq. (6),

$$Z^{\mathrm{FTLM}} \approx \frac{\mathcal{D}}{R} \sum_{r=1}^{R} \sum_{k=1}^{N_L} w_k^{(r)} e^{-\beta \theta_k^{(r)}} . \tag{6}$$

If symmetries are resolved, the same construction is applied separately in each symmetry sector $\Gamma$, Eq. (7),

$$Z^{\mathrm{FTLM}} \approx \sum_\Gamma \frac{\mathcal{D}_\Gamma}{R} \sum_{r=1}^{R} \sum_{k=1}^{N_L} w_k^{(\Gamma, r)} e^{-\beta \theta_k^{(\Gamma, r)}} . \tag{7}$$

In the present work, we only use $S^z$ symmetry, so the sectors are labeled by the magnetization quantum number $M$.

The observables considered in the precision analysis of section 3.1 are the heat capacity and the zero-field magnetic susceptibility. In units of $k_{\mathrm{B}} = 1$, the heat capacity is obtained from the energy variance, $C = \beta^2(\langle H^2\rangle - \langle H\rangle^2)$, where the thermal moments are computed from Eq. (8)

$$\langle H^p\rangle^{\mathrm{FTLM}} \approx \frac{1}{Z^{\mathrm{FTLM}}}\sum_\Gamma \frac{\mathcal{D}_\Gamma}{R}\sum_{r=1}^{R}\sum_{k=1}^{N_L} w_k^{(\Gamma,r)}\left[\theta_k^{(\Gamma,r)}\right]^p e^{-\beta\theta_k^{(\Gamma,r)}} \tag{8}$$

In zero-field, the susceptibility (in units $k_{\mathrm{B}} = 1$, $\mu_{\mathrm{B}} = 1$), is evaluated from the sector partition functions (for plots of $\chi$ in section 3.1, we set the gyromagnetic factor to $g = 2$), Eq. (9),

$$\chi = g^2\beta\langle(S^z)^2\rangle = g^2\beta\,\frac{\sum_M M^2 Z_M}{\sum_M Z_M}\,. \tag{9}$$

In the Results section below, FP32 refers to the single-precision Lanczos recursion on the GPU, whereas FP64 refers to the double-precision CPU reference calculation. In both cases, the small tridiagonal eigenproblem is solved on the host in FP64.

**Matrix-free GPU kernel.** The memory required for an explicitly stored Hamiltonian grows rapidly with the sector dimension $\mathcal{D}_M(N,s)$ (an analytical expression was derived in Ref. [30]). For $N$ sites with local spin $s$ and $N_{\mathrm{B}}$ pairwise interactions, the number $N_{\mathrm{nz}}$ of nonzero entries in the $M$-sector Hamiltonian matrix in a compressed sparse (CS) representation is

$$N_{\mathrm{nz}} = \mathcal{D}_M(N,s) + 2N_{\mathrm{B}}\sum_{a,b}\mathcal{D}_{M-a-b}(N-2,s)\,, \tag{10}$$

where we use the convention that $\mathcal{D}_{M'}(N-2,s) = 0$ if $M'$ lies outside the allowed magnetization range. The first term in Eq. (10) accounts for the diagonal contribution (from the $s_i^z s_j^z$ terms). The second term counts the off-diagonal contributions generated by the ladder-operator terms $s_i^+ s_j^-$ and $s_i^- s_j^+$. For $s_i^+ s_j^-$, the source-state quantum numbers satisfy $a \in \{-s,\dots,s-1\}$ and $b \in \{-s+1,\dots,+s\}$; the ranges are interchanged for $s_i^- s_j^+$, giving the factor of two. As an example, for the $M = 0$ sector of the icosahedron ($N_{\mathrm{B}} = 30$) the memory required by a double-precision CS representation grows from 35 MB at $s = 1$ to 587 GB at $s = 3$ (Table 1). Thus, explicit storage of the Hamiltonian quickly becomes impractical on workstation GPUs, motivating a matrix-free evaluation of the Hamiltonian action.

Table 1: Memory required to store the Heisenberg Hamiltonian in a MATLAB-style 64-bit sparse format[a] for the $M = 0$ sector of the icosahedron ($N = 12$, $N_B = 30$) for different local spins $s$.

| $s$ | $\mathcal{D}_0$ | $N_{\mathrm{nz}}$ | Memory |
|---|---|---|---|
| 1/2 | 924 | 16 044 | 0.26 MB |
| 1 | 73 789 | 2 150 149 | 35 MB |
| 3/2 | 1 703 636 | 61 539 956 | 998 MB |
| 2 | 19 611 175 | 797 089 975 | 12.9 GB |
| 5/2 | 144 840 476 | 6 342 881 276 | 103 GB |
| 3 | 786 588 243 | 36 264 501 483 | 587 GB |

[a] MATLAB stores sparse matrices in compressed-column 64-bit sparse format (8 B per double-precision value, 8 B per row index, and 8 B per column pointer, giving $16N_{nz} + 8(\mathcal{D}_M + 1)$ bytes).

For this reason, we evaluate the SpMV directly, without storing the Hamiltonian matrix [31]. We use the row-wise gather formulation of Schnack et al. [16], in which each output element is computed independently. In this formulation, the outer loop runs over the row index $i$, and the corresponding uncoupled basis state $|i\rangle$ is used to generate all connected source states $|j\rangle$ for which $\langle i|H|j\rangle \neq 0$. The contribution to the output vector is then accumulated as $y_i = \sum_j \langle i|H|j\rangle x_j$. The connected states are generated on the fly rather than being read from a stored sparse matrix, and their positions in the $M$-sector basis are obtained through a state-to-index map, for which the lookup strategies discussed below are used. Because each thread writes to exactly one output element, this gather formulation avoids write conflicts. This property is useful for shared-memory CPU parallelism and becomes particularly important on GPUs, where a scatter formulation would lead to many concurrent atomic additions to the same output vector.

To make this matrix-free gather formulation explicit, we first describe the integer encoding of product-basis states and then the per-thread evaluation of one output-vector element. Each basis state is encoded by an integer, Eq. (11),

$$n = \sum_{k=0}^{N-1} (m_k + s) d^k \ , \tag{11}$$

where $d = 2s + 1$, so that the full Hilbert space has dimension $\mathcal{D} = d^N$, and the sites are labeled $k = 0,1, \dots, N - 1$. The shifted local quantum number $a_k = m_k + s$ is the $k$-th base-$d$ digit of $n$ and can be obtained by integer division and modulo operations, $a_k = \lfloor n \ / \ d^k \rfloor \bmod d$. For each $M$-sector, a simple implementation stores a basis array containing the integer labels of the sector basis states. In addition, a state-to-index map is required to locate a generated state label within the sector basis.

Each GPU thread is assigned one sector-basis index $t$. It reads the corresponding state label $n_t = \text{basis}[t]$, decomposes it into the local quantum numbers $\{m_k\}$, and accumulates the corresponding output-vector element $w[t]$ from the input vector $v$, Eq. (12),

$$w[t] \;=\; J \sum_{\langle i,j\rangle} m_i\, m_j\, v[t] \;+\; \sum_{\langle i,j\rangle} \left[ c_{ij}^{+} v \left[\text{lookup}\left[n_{ij}^{+}\right]\right] \;+\; c_{ij}^{-} v \left[\text{lookup}\left[n_{ij}^{-}\right]\right] \right]. \tag{12}$$

Here $n_{ij}^{+}$ and $n_{ij}^{-}$ denote the source-state labels connected to $n_t$ by the two ladder-operator contributions on bond $\langle i,j\rangle$, and $c_{ij}^{+}$ and $c_{ij}^{-}$ are the corresponding matrix elements (Eqs. 13 and 14, respectively). These coefficients and source labels are generated directly from the digits of $n_t$. The lookup operation then converts each valid source-state label into its sector-basis index.

$$c_{ij}^{+} \;=\; \frac{J}{2} \sqrt{s(s+1) \;-\; m_i(m_i - 1)} \sqrt{s(s+1) \;-\; m_j\left(m_j + 1\right)} \tag{13}$$

$$c_{ij}^{-} \;=\; \frac{J}{2} \sqrt{s(s+1) \;-\; m_i(m_i + 1)} \sqrt{s(s+1) \;-\; m_j\left(m_j - 1\right)} \tag{14}$$

**State-to-index mapping.** The matrix-free kernel requires a map from a generated product-state label to its position in the *M*-sector basis. Different mapping strategies realize different trade-offs between memory footprint and lookup cost. The simplest option is a full lookup table, which stores one integer per label in the full Hilbert space of size $\mathcal{D} = d^N$. The stored value is the corresponding sector index, while labels outside the sector are marked as invalid. This gives direct access with a single table read but requires $4\mathcal{D}$ bytes when 32-bit indices are sufficient, i.e., for sector dimensions $\mathcal{D}_M < 2^{32}$. For $N = 12$, this amounts to 977 MB at $s = 2$ and 8.7 GB at $s = 5/2$. Most of this memory is allocated to labels that do not belong to the sector basis.

Several lower-memory alternatives have therefore been developed. Hash-based searches store only sector states but introduce data-dependent memory access and search costs [32]. Lin's two-dimensional search [33] reduces the table size by partitioning the system into two halves, with rank tables of size approximately $d^{N/2}$ for each half. In our GPU-adapted variant of this decomposition, the sector basis is grouped by the sub-magnetization of one half. The sector index is then recovered from a fixed-length digit sum over one half of the system, two reads from the half-system rank tables, and two reads from a small offset table stored in constant memory. This gives a data-independent lookup pattern for a fixed system size, at the price of additional arithmetic compared with a full lookup table. In our single-GPU benchmarks the resulting Lin kernel runs at roughly half the throughput of the CLT-based variant.

Alternatively, the combinatorial-ranking (CR) strategy of Schnack et al. [16] computes the sector index of a basis state directly from its digits, without storing a per-label index structure. We denote by $D(m, A)$ the number of length-$m$ digit strings with digits in $\{0, \dots, 2s\}$ and total digit sum $A$. These dimensions obey the recursion Eq. (15),

$$D(m, A) = \sum_{q=0}^{2s} D(m - 1, A - q), \qquad D(0,0) = 1, \tag{15}$$

with $D(m, A) = 0$ outside the allowed range. For ranking in ascending integer-label order, it is useful to precompute the cumulative table Eq. (16),

$$D_{\mathrm{c}}(m, A, a) = \sum_{q=0}^{a-1} D(m, A - q), \qquad a = 0, \dots, 2s. \tag{16}$$

The GPU-aware reformulation rests on three choices. First, the cumulative table replaces the original algorithm's data-dependent inner scan by a single table lookup at each digit position. Second, $D_{\mathrm{c}}$ is staged in shared memory rather than read from constant memory during the lookup. This avoids the broadcast-style serialization of constant memory for thread-divergent indices and allows the table layout to be chosen to minimize shared-memory bank conflicts. Third, both the ranking step, from state label to sector index, and the unranking step, from sector index to state label, traverse the $N$ digit positions in fixed-length loops. The per-digit work is therefore independent of the input state, so that all threads in a warp follow the same instruction stream. The outer ranking structure is unchanged from Ref. [16]; the GPU adaptation consists of the three described reformulations of the inner accumulation and memory layout.

For the digit convention of Eq. (11), the total digit sum in an $M$-sector is Eq. (17),

$$A_{\mathrm{tot}} = Ns + M \,. \tag{17}$$

Algorithm 2 summarizes the ranking step. The inverse map uses the same outer-loop structure and determines each digit by comparing the remaining rank with the cumulative counts. In the row-wise gather formulation of the SpMV, the GPU thread assigned to output index $t$ first reconstructs the corresponding state label via the inverse map and then ranks each connected source label back to its sector index.

**Algorithm 1.** GPU-adapted combinatorial ranking for a state label $x$.

**Input:** integer label $x$ of the basis state; cumulative dimension table $D_c(m, A, a)$ in shared memory; total digit sum $A_{tot} = Ns + M$ of the sector.

1. Initialize rank $\leftarrow$ 0 and $A \leftarrow A_{tot}$.
2. For $p = N - 1, \dots, 0$: read the $p$-th digit $a_p$ of $x$, then update rank $\leftarrow$ rank $+ D_c(p, A, a_p)$ and $A \leftarrow A - a_p$.
3. Return rank.

The auxiliary storage of the GPU-adapted ranking scheme consists of the $D_c$ table, which occupies at most a few ten kilobytes for the systems considered here, together with the bond list and the small coefficient tables. No basis array and no per-label index structure are stored. For the $s = 1/2$ icosidodecahedron with $N = 30$, this saves approximately one GB of device memory relative to the CLT-based implementation, where both the basis array and the CLT itself are stored.

**Compressed lookup table (CLT).** The CLT provides a stored-map alternative to CR. It keeps the state-to-index translation data-independent and $\mathcal{O}(1)$, but reduces the memory footprint of a full lookup table by a factor of 16. The idea is to store the occupancy of the full label space as a bit vector and to augment it with block-wise prefix counts, so that the sector index of a valid label can be recovered by a rank query on this bit vector [34].

The label space $\{0, \dots, \mathcal{D} - 1\}$ is partitioned into consecutive blocks of 32 labels. For each block $b$, two 32-bit quantities are stored. The first is an occupancy mask $M_b$, whose bit $j$ is set if and only if the label $32b + j$ belongs to the sector. The second is a prefix count $C_b$, defined as the number of in-sector labels in all preceding blocks, Eq. (18),

$$C_b = \sum_{\ell=0}^{b-1} \mathrm{popcount}(M_\ell). \tag{18}$$

This construction assumes that the sector basis is ordered by increasing integer label. Under this convention, the number of set bits before a given bit position within the block is exactly the local rank of the corresponding label inside that block.

Given a generated state label $n$, the block index and bit position are obtained from Eq. (19),

$$b = \lfloor n/32 \rfloor, \quad j = n \bmod 32 . \tag{19}$$

The occupancy bit of $M_b$ determines whether $n$ belongs to the sector. If it does, the sector index is Eq. (20),

$$\mathrm{lookup}(n) = C_b + \mathrm{popcount}(M_b \,\&\, (2^j - 1)) \,, \tag{20}$$

where the mask $2^j - 1$ selects all bits below position $j$. (In the Hamiltonian application, the occupancy test succeeds for valid ladder-operator moves because $M$ is conserved.) The lookup therefore consists of two table reads, a bit test, and one population count. On NVIDIA GPUs, these operations map to efficient integer instructions, including `POPC` for the population count.

**Algorithm 2.** CLT lookup for a generated state label $n$.

**Input:** state label $n$; per-block occupancy masks $M_b$; per-block prefix counts $C_b$.

1. Compute $b = \lfloor n/32 \rfloor$ and $j = n \bmod 32$.
2. If bit $j$ of $M_b$ is zero, return $-1$ ($n$ does not belong to the sector).
3. Return $C_b + \mathrm{popcount}(M_b \,\&\, (2^j - 1))$.

As a simple example, consider a block $b$ in which the labels $32b + 0, 32b + 2, 32b + 4$, and $32b + 5$ belong to the sector. The occupancy mask is then $M_b = 00110101_2$. To look up the label $32b + 4$, one has $j = 4$. The lower-bit mask $2^4 - 1 = 01111_2$ selects bits 0 through 3, and $M_b \,\&\, 01111_2 = 00101_2$ has population count 2. The returned index is therefore $C_b + 2$, which is the position of $32b + 4$ after the two earlier in-sector labels in the same block.

The storage cost is 8 bytes per block, namely 4 bytes for $M_b$ and 4 bytes for $C_b$. Since each block represents 32 labels, the CLT requires Eq. (21),

$$8\,\lceil \mathcal{D}/32 \rceil \approx \mathcal{D}/4 \,, \tag{21}$$

bytes, compared with $4\mathcal{D}$ bytes for a full 32-bit lookup table. The factor-of-16 reduction follows directly from this ratio. Empty blocks have $M_b = 0$; they contribute to the prefix counts but are not accessed during matrix-vector products generated within an $M$-sector. The final block is padded with zero bits if $\mathcal{D}$ is not a multiple of 32.

The CLT is constructed in a single pass over the sorted sector basis. For each basis label $n$, the corresponding bit $n \bmod 32$ is set in $M_{\lfloor n/32 \rfloor}$. The prefix counts $C_b$ are then obtained by a cumulative sum of the block population counts. In the implementation, the basis array and the CLT reside in global memory, while the bond list and precomputed powers $d^k$ are stored in constant memory. For the $N = 12, s = 2$ icosahedron, the CLT occupies 61 MB instead of 977

MB for the full lookup table. For the $N = 30, s = 1/2$ icosidodecahedron, it requires 268 MB instead of 4.3 GB.

Compared with CR, the CLT stores an explicit compressed representation of the state-to-index map. It therefore uses more memory but replaces the digit-by-digit rank reconstruction by a compact, fixed lookup sequence. The two approaches thus occupy different points in the same memory-throughput design space: the CLT favors per-lookup throughput when the table fits in device memory, whereas GPU-adapted CR minimizes memory use and would represent the natural fallback whenever the lookup table consumes a significant share of the available memory, for example, in multi-GPU runs in which the lookup table must be replicated on every device while the Lanczos vectors are sliced; with its $D_\mathrm{c}$-table of a few ten kilobytes, CR shifts the memory bottleneck back to the Lanczos vectors.

Table 2: Memory comparison (in MB) for the $s = 2$ icosahedron, $M = 0$ sector. Floating-point quantities (Hamiltonian and three Lanczos work vectors) are reported in FP64. Index-valued components are stored as int32.[a]

| Component | CS | Full Lookup | CLT |
|---|---|---|---|
| Hamiltonian | 12 910 | — | — |
| Lookup table | — | 977 | 61 |
| Basis array | — | 78 | 78 |
| Work vectors (3×) | 471 | 471 | 471 |
| Sum | ~13 381 | ~1526 | ~610 |

[a]In an FP32 realization of the Krylov iteration, the work vectors occupy only 235 MB.

**Computational setup.** All calculations were performed on a workstation equipped with an NVIDIA RTX 4000 SFF Ada Generation GPU with 20 GB of GDDR6 memory and a nominal memory bandwidth of 280 GB/s, and an Intel Core Ultra 9 285T CPU with 24 cores and 64 GB of dual-channel DDR5 memory. CUDA kernels were compiled with `MATLAB R2025b`'s mexcuda, and the OpenMP-parallelized CPU kernels were compiled with MSVC and OpenMP.

Large-scale CPU implementations such as `spinpack` distribute the Hilbert space and the associated vectors across MPI ranks, while the matrix-free SpMV within each rank is parallelized over shared-memory threads using the row-wise loop rearrangement of Ref. [16]. The present implementation instead targets a single workstation GPU. In the CLT-based variant, the sector basis and lookup structure reside in GPU global memory, whereas no per-label lookup structure is stored in the CR variant. In both cases, the SpMV is parallelized over output elements, assigning one GPU thread to each sector-basis state. This single-device approach avoids distributed-memory communication during the SpMV but is limited by GPU memory capacity. The performance is therefore determined primarily by memory traffic associated with the input vector, basis representation, and state-to-index mapping.

For the CPU reference calculations, we use a batched Lanczos variant that processes $B = 8$ independent Krylov chains simultaneously [35]. Because the Hamiltonian row structure (digit decomposition, bond iteration, source-state generation, and state-to-index lookup) depends only on the output basis state and not on the input vector, this per-row work can be shared across the $B$ input vectors in every Lanczos step. The lookup and digit-decomposition costs are therefore amortized over $B$ SpMV operations, improving the ratio of arithmetic work to memory traffic for the off-diagonal spin-flip terms. An alternative would be to distribute independent Lanczos chains across workers using `MATLAB`'s `parfor` construct. This can achieve similar wall-clock times for large numbers of random vectors, but it does not exploit the cache-level reuse of the batched approach. We therefore use the batched Lanczos variant for all CPU timings, because it is the most efficient single-node CPU implementation available to us for the workloads considered here.

## 3. Results

We first present empirical FP32-vs-FP64 comparisons for representative FTLM calculations and show that the observed differences remain negligible compared with the stochastic FTLM uncertainty. The stronger loss of Krylov-space orthogonality in FP32 produces more ghost eigenvalues, but these mainly redistribute spectral weight within already converged Ritz clusters. Finally, a heuristic backward-error argument explains that the relevant error scale is controlled primarily by the product of the unit roundoff $u$ and the spectral bandwidth $W$, rather than showing any systematic growth with the Hilbert-space dimension.

**3.1 Precision analysis.** The adequacy of FP32 arithmetic for FTLM calculations is assessed by comparing two scales: (i) the empirical standard deviation $\sigma_{\mathrm{emp}}(O)$ of a thermodynamic observable $O$ with respect to independent FTLM calculations, each built from $R$ random starting vectors per magnetization sector, and (ii) the observed deviation $|\Delta O| = |O_{\mathrm{FP32}} - O_{\mathrm{FP64}}|$. FTLM calculations in FP32 and FP64 share the same random seeds, so that differences in results reflect the combined effect of FP32 arithmetic and the GPU-vs-CPU floating-point reduction order. The analysis of section 3.3 treats both effects as contributions to an effective operator-level perturbation. Thus, while the comparison should not be interpreted as isolating mantissa precision alone, it remains the relevant test of the GPU implementation.

Unless noted otherwise, for the following examples, we choose $R = 100$ random vectors per $M$-sector for $N_L = 100$ Lanczos steps each. As a consistency check, we confirmed in all cases that the lowest Ritz values from the FP32 GPU runs agreed with the FP64 CPU references at the expected single-precision roundoff level.

Figure 1 shows for the icosahedron with either $s = 1$ or $s = \frac{3}{2}$ sites that FP32 and FP64 results for the heat capacity $C$ and the magnetic susceptibility $\chi$ coincide at graphical resolution across the full $T$ range. The absolute FP32–FP64 deviations, $|\Delta C|$ and $|\Delta\chi|$, are plotted on the right logarithmic axis of each panel and remain six to seven orders of magnitude below the observable across the entire temperature range, which provides evidence that FP32 suffices for these properties.

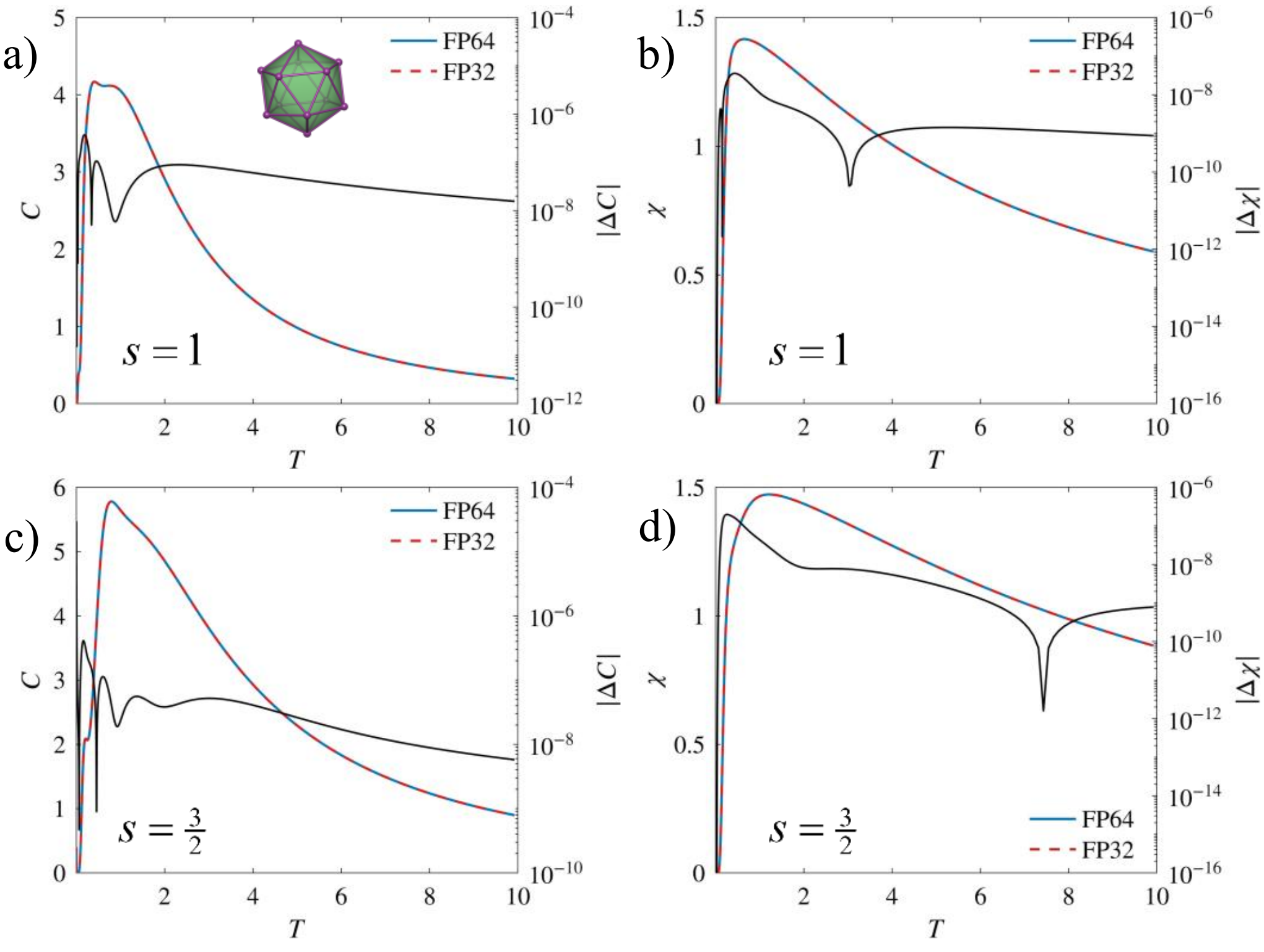

Figure 1: Thermodynamic observables and precision error for the antiferromagnetic icosahedron with local spin $s = 1$ (panels a, b) or $s = 3/2$ (c, d); left: heat capacity $C$, right: magnetic susceptibility $\chi$. FP64 (solid blue) and FP32 (dashed red) are plotted on the linear left $y$-axis; both types of curves coincide at the resolution of the plot. The pointwise precision error, $|\Delta C| = |C_{\mathrm{FP32}} - C_{\mathrm{FP64}}|$ or $|\Delta\chi| = |\chi_{\mathrm{FP32}} - \chi_{\mathrm{FP64}}|$, is plotted in black against its logarithmic right $y$-axis.

To compare this accuracy to the inherent stochastic FTLM uncertainty, we consider the standard deviation $\sigma_{\mathrm{emp}}$ with respect to FTLM results based on $N_s$ distinct random seeds [13], Eq. (22),

$$\sigma_{\mathrm{emp}}(O) = \sqrt{\frac{1}{N_s}\sum_{s=1}^{N_s}(\langle O\rangle_s - \bar{O})^2} \quad (22)$$

where $\langle O\rangle_s$ denotes the FTLM estimate for the observable at seed $s$, and $\bar{O} = \frac{1}{N_s}\sum_{s=1}^{N_s}\langle O\rangle_s$. We ran $N_s = 50$ statistically independent calculations, each with its own set of $R = 100$ random vectors. Pooling all 50 runs yields an effective FTLM sample size of $R_{\mathrm{eff}} = 5000$. Figure 2 presents $\sigma_{\mathrm{emp}}$ for the heat capacity and susceptibility for both icosahedra ($s = 1$ and $s = \frac{3}{2}$) alongside an approximate theoretical bound $\sigma_{\mathrm{theo}} = O/\sqrt{Z_{\mathrm{eff}}}$ (see [13] and references cited therein) for the pooled $R_{\mathrm{eff}} = 5000$ FTLM calculation ($Z_{\mathrm{eff}} = \mathrm{Tr}(e^{-\beta(H-E_0)})$ is the effective partition function, with $E_0$ being the global ground-state energy). The precision error of Figure 1 lies four to five orders of magnitude below both stochastic scales across the entire temperature range. As the statistical error approximately falls as $1/\sqrt{R}$, the break-even point $R^*$ at which $\sigma_{\mathrm{emp}}$ equals $|\Delta_{\mathrm{FP32}}|$ would lie beyond any practical application. The FP32 precision therefore does not become the limiting uncertainty for thermodynamic observables.

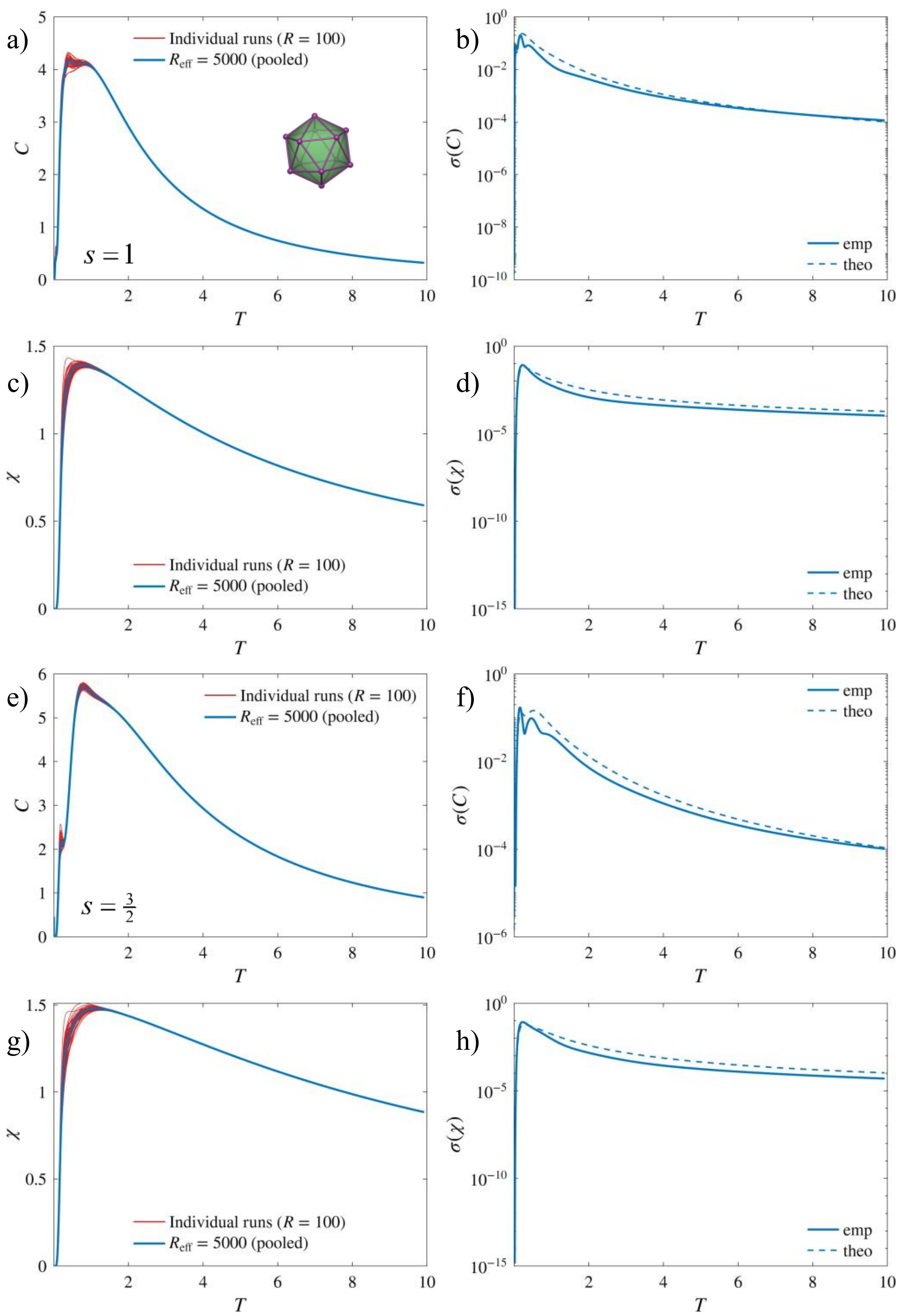

Figure 2: Multi-seed stochastic error analysis for the heat capacity and magnetic susceptibility of the icosahedron with $s = 1$ (a, b, c, d) or $s = 3/2$ (e, f, g, h). Left: 50 independent FTLM estimates (red curves, $R = 100$ each) and the pooled $R_{\text{eff}} = 5000$ results (blue). Right: empirical and theoretical standard deviations across the 50 runs. See main text for details.

We additionally checked the agreement between FP32 and FP64 results on four additional systems chosen to span different lattice topologies (rings and polyhedra), different spin values and a broad range of Hilbert-space dimensions, up to the $s = 1/2$ icosidodecahedron with ~$1.6 \cdot 10^8$ states in the $M = 0$ sector, which is near the upper limit at which an FP64 reference is still tractable on a workstation. In all cases FP32 and FP64 are practically indistinguishable at graphical resolution (Figure 3), in line with the dimension-independence suggested by arguments given in section 3.3 below.

The next two subsections address why the larger rate of ghost eigenvalues in FP32 does not affect the small FP32-vs-FP64 difference and explain that the error scale is controlled by the product of the unit roundoff and the spectral bandwidth rather than by the Hilbert-space dimension.

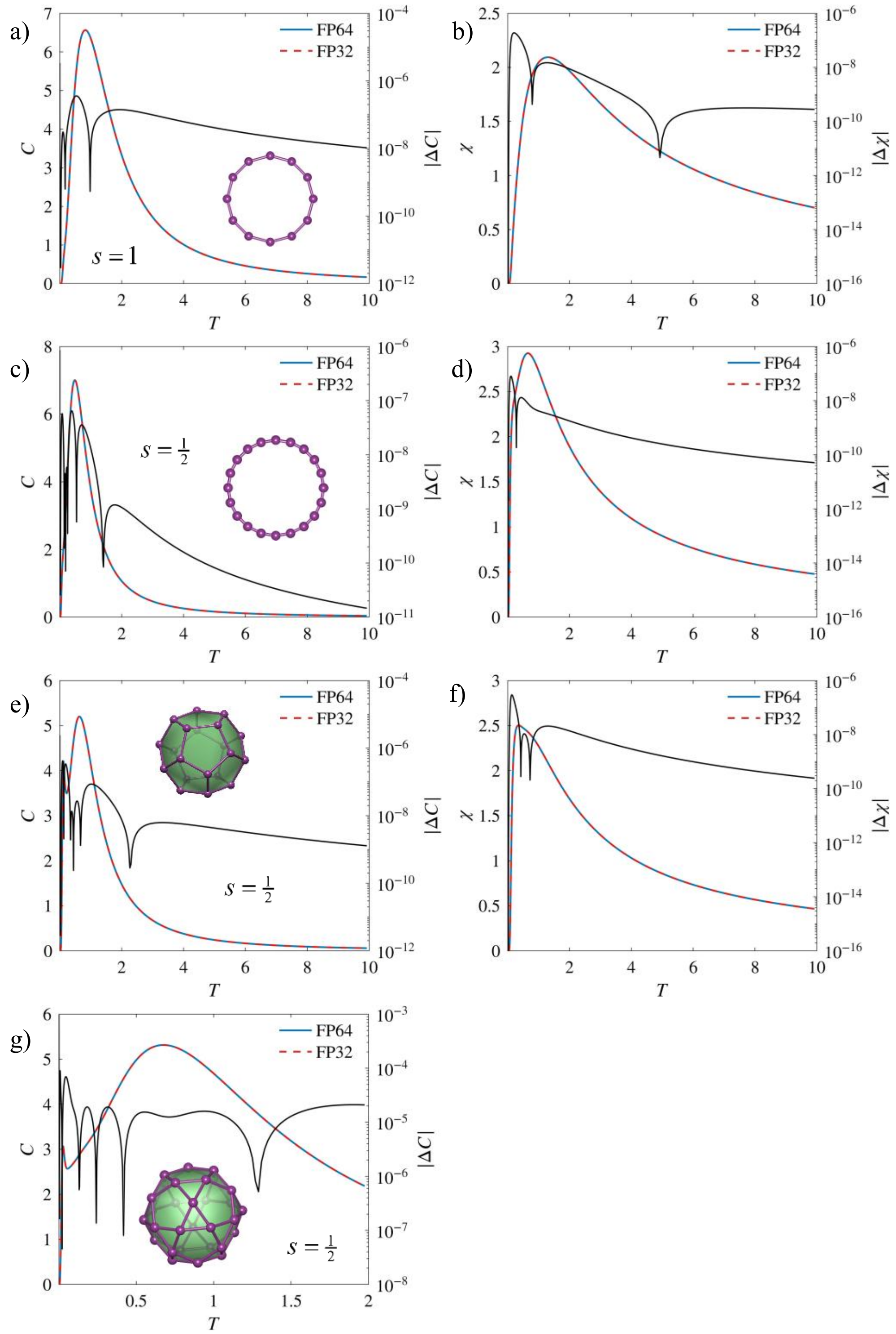

Figure 3: Thermodynamic observables and precision error (for details, see caption to Figure 1) for the $s = 1$, $N = 12$ Heisenberg ring (panels a, b), the $s = 1/2$ , $N = 20$ ring (c, d), the $s = 1/2$ dodecahedron (e, f) and the $s = 1/2$ icosidodecahedron (g). For the icosidodecahedron the calculation was restricted to the $M = 0$ sector (thus $\chi = 0$, not shown) and only $R = 8$ random vectors were used.

**3.2 Ghost eigenvalues and spectral weight.** Here we address a possible concern associated with finite-precision Lanczos calculations without reorthogonalization: FP32 produces more ghost eigenvalues than FP64. When a Ritz pair has converged but the Lanczos vectors are no longer fully orthogonal to the associated eigendirection, roundoff components along this direction can re-enter the recursion. The same physical eigenvalue may then reappear as a near-duplicate, or ghost, Ritz value. This effect is stronger in FP32 because the roundoff level is larger.

For FTLM, however, the relevant object is not the list of Ritz values alone, but the Ritz values together with their spectral weights. Changes in individual $\theta_k$, or a splitting of the spectral weights into near-duplicate values, matter only through their effect on such weighted sums, cf. Eqs. (4) and (5) above. These sums can be written compactly in terms of a cumulative spectral weight function, Eqs. (23) and (24),

$$F(E) = \sum_{k:\theta_k \leq E} w_k \qquad (23)$$

for example,

$$Z_r \approx \sum_k w_k e^{-\beta\theta_k} = \int e^{-\beta E}\, \mathrm{d}F(E) \qquad (24)$$

This representation clarifies what should be compared between FP32 and FP64, namely whether the accumulated spectral weight is stable on the energy scale relevant for the thermal sums. The pointwise difference between two step functions is not by itself a useful measure of the FTLM error, since small horizontal shifts of Ritz values can produce visible vertical differences at the jumps.

We identify near-duplicate Ritz values using the Cullum–Willoughby criterion [36]. Let $T^{(1)}$ be the tridiagonal matrix (eigenvalues $\theta_j^{(1)}$) obtained by deleting the first row and column of $T$. An eigenvalue $\theta_k$ of $T$ is flagged as a ghost if the deflation distance, defined in Eq. (25),

$$\ell_k = \min_j \left|\theta_k - \theta_j^{(1)}\right|, \qquad (25)$$

falls below a tolerance, $\ell_k < \tau$. We choose a common tolerance for FP32 and FP64 based on the FP32 backward-error scale (cf. section 3.3), $\tau = C_\tau\, u_{FP32}\, W_T$, where $C_\tau$ is a numerical factor of order unity, $u_{\mathrm{FP32}}$ is the FP32 unit roundoff, and $W_T = \theta_{\max} - \theta_{\min}$. This threshold is large enough to identify FP32 ghosts while still being applied identically to both precisions. The results should therefore be interpreted as diagnostic counts within this fixed tolerance, not as absolute precision-independent ghost rates.

Figure 4 provides a visualization for the $M = 0$ sector in the $s = 1$ icosahedron. For each $\theta_k$ of the FP64 (left) or FP32 (right) Lanczos run (both based on the same random seed), the associated $\ell_k$ is shown on a logarithmic scale. Regular eigenvalues, i.e., values not flagged as ghosts, appear as black circles. Ghosts (red circles) are more numerous in FP32 (10) than in FP64 (4).

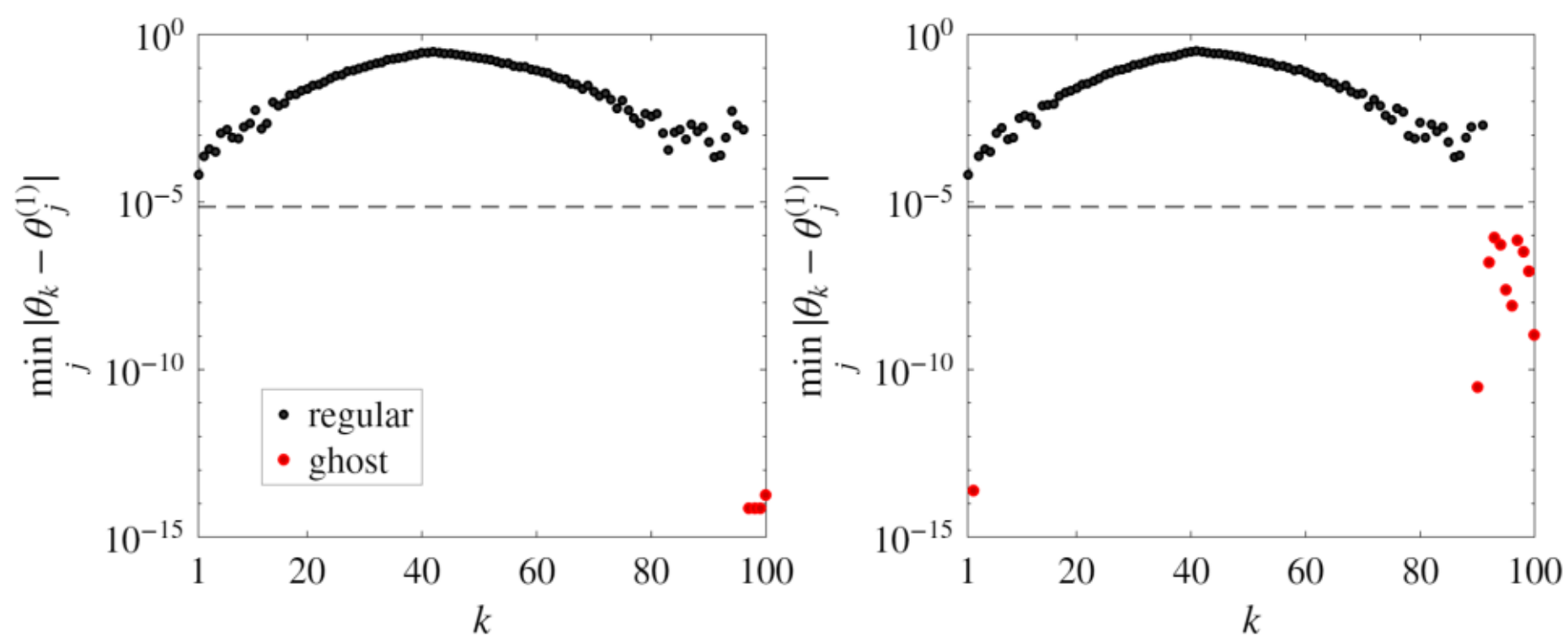

Figure 4: Ghost diagnostic for the Lanczos Ritz spectrum – based on the same random starting vector for both FP64 (left) and FP32 (right) – in the $M = 0$ sector of the $s = 1$ icosahedron. Black circles: regular eigenvalues; red circles: ghosts in the Cullum–Willoughby sense.

Note that a ghost-flagged Ritz value may correspond to a genuine physical eigenvalue that is represented more than once due to loss of orthogonality. In the standard interpretation, such ghosts are repeated representations of spectral information that has already converged. The meaningful quantity is therefore not the weight of an individual cluster member, but the total cluster weight. Figure 5 places values $\theta_k$ and $\theta_l$ in the same cluster whenever $|\theta_k - \theta_l| \leq \tau$. Except for one FP32 ghost ($\theta_2 \approx -18.53$, $w_2 \approx 0$) all ghosts (red circles) form clusters. The summed total weights of the latter (marked with blue circles) show only small deviations between FP64 and FP32. Thus, the additional FP32 ghosts mainly redistribute spectral weight within already converged Ritz clusters rather than creating independent spectral content. This explains why the larger FP32 ghost count is not in itself a threat to the FTLM observables.

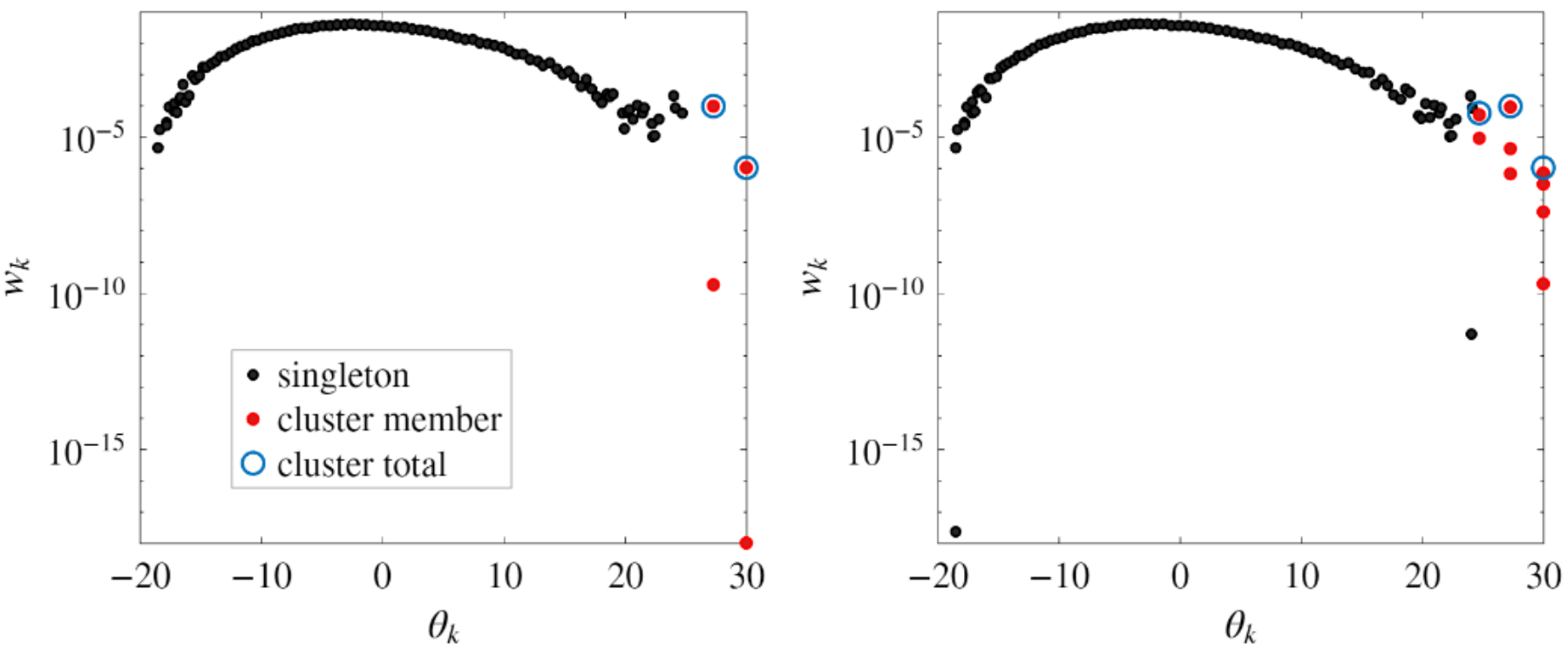

Figure 5: Cluster diagnostic for the Krylov-space energy spectrum in the $M = 0$ sector of the $s = 1$ icosahedron. The weights $w_k$ are plotted on a logarithmic scale versus eigenvalues $\theta_k$ (in ascending order). Black circles: singletons (not flagged by the ghost diagnostic, or not assigned to a cluster), red circles: ghost-flagged Ritz values, blue circles: total weight of clusters containing ghost-flagged near-duplicate eigenvalues.

**3.3 FP32 error analysis.** We now estimate the scale of the FP32-vs-FP64 changes in the thermodynamic sums from a backward-error point of view. The finite-precision Lanczos recursion on a Hamiltonian $H$ admits a backward-error interpretation [37–42]. In this view, the computed tridiagonal matrix can be interpreted as the result of an exact Lanczos process for a nearby, effectively perturbed spectral problem. The perturbation arises from rounding errors in the SpMV, dot products and three-term recurrence. Up to implementation-dependent numerical factors, its operator scale, $\|\delta H\| = \mathcal{O}(u\,W)$, is set by the unit roundoff $u$ ($u_{\mathrm{FP32}} \approx 6 \times 10^{-8}$) and the spectral bandwidth $W = E_{\max} - E_{\min}$ of $H$, rather than by the Hilbert-space dimension. We write $H_\lambda = H + \lambda\,\delta H$, $0 \le \lambda \le 1$, and denote by $\delta Q = \frac{dQ(\lambda)}{d\lambda}\Big|_{\lambda=0}$ the first-order change of a thermodynamic scalar $Q$ along this interpolation. The corresponding FP32-vs-FP64 deviation observed in the numerical data is denoted by $\Delta Q$. For energy-like thermodynamic quantities, the backward-error picture gives the scale estimate $|\Delta Q| \sim |\delta Q| = \mathcal{O}(u\,W)$. For the Heisenberg clusters considered here, with $J = 1$ and $W = \mathcal{O}(10)$, this gives $u_{\mathrm{FP32}} W \sim 10^{-6}$, much smaller than the relevant thermodynamic energy scales and well below the stochastic uncertainty $\sigma_{\mathrm{emp}}$ arising from the finite number of random starting vectors.

The susceptibility, $\chi = \beta\,\langle (S^z)^2 \rangle$, deserves a separate comment because it contains an explicit factor of $\beta$. In zero field, we evaluate $\langle (S^z)^2 \rangle$ from the $M$-sector partition functions $Z_M$. Under an effective backward perturbation $\delta H$, the first-order change in $Z_M$ is $\delta Z_M / Z_M = -\beta\,\langle \delta H \rangle_M$, where $\langle \delta H \rangle_M = \mathrm{Tr}_M\big(\mathrm{e}^{-\beta H}\,\delta H\big)/Z_M$. Equivalently, varying the normalized sector weights $p_M = Z_M / Z$ gives Eq. (27),

$$\delta\langle (S^z)^2 \rangle = -\beta\,\mathrm{Cov}_M\big(M^2, \langle \delta H \rangle_M\big)\,, \tag{27}$$

where the covariance ($\mathrm{Cov}(X,Y) \equiv \langle X\, Y\rangle - \langle X\rangle \langle Y\rangle$) is taken over the sector weights $p_M$. Applying the Cauchy–Schwarz inequality, $|\mathrm{Cov}(X,Y)| \le \sigma(X)\, \sigma(Y)$, where $\sigma(X) \equiv \sqrt{\langle X^2\rangle - \langle X\rangle^2}$, and using $\sigma_M(\langle \delta H\rangle_M) \lesssim \|\delta H\|$, we obtain $|\delta\langle (S^z)^2\rangle| \lesssim \beta\, \sigma_M(M^2)\, \varepsilon_{\mathrm{FP32}}\, \|\delta H\| \sim \beta\, \sigma_M(M^2)\, u\, W$. This directly gives the susceptibility scale $|\Delta\chi| \sim |\delta\chi| \lesssim \beta^2\, \sigma_M(M^2)\, u\, W$. For the antiferromagnetic systems considered here, with gapped, non-magnetic ground states, $\sigma(M^2)$ vanishes exponentially in $\beta\, \Delta_{\mathrm{gap}}$. Hence the product $\beta^2 \sigma_M(M^2)$ remains bounded. This is consistent with the observed absence of any low-temperature precision loss (section 3.1).

The heat capacity, $C = \beta^2\, (\langle H^2\rangle - \langle H\rangle^2) = \beta^2 \mathrm{Var}(H)$, also requires a separate comment because of the explicit factor $\beta^2$. In the backward-error picture, $\langle H^2\rangle$ and $\langle H\rangle$ originate from the same effective perturbed Hamiltonian $H + \delta H$. To obtain a correlated estimate, we consider $C(\lambda) = \beta^2\, \left(\langle H_\lambda^2\rangle_\lambda - \langle H_\lambda\rangle_\lambda^2\right)$, where the thermal averages $\langle\cdot\rangle_\lambda$ are taken with respect to $H_\lambda$. To first order in $\lambda$, changes of $C$ come from two correlated effects: the explicit change of the energy moments and the corresponding change of the Boltzmann weights. In this first-order spectral perturbation picture, we use Eq. (28),

$$\frac{d}{d\lambda}\langle A_\lambda\rangle_\lambda = \langle \frac{dA_\lambda}{d\lambda}\rangle_\lambda - \beta \mathrm{Cov}_\lambda(A_\lambda, \delta H)\,, \tag{28}$$

which yields

$$\delta C \;\approx\; 2\,\beta^2\, \mathrm{Cov}(H, \delta H) \;-\; \beta^3\, \kappa_{2,1}\,, \tag{29}$$

with Eq. (30),

$$\kappa_{2,1} \;= \langle (H - \langle H\rangle)^2 (\delta H - \langle \delta H\rangle)\rangle = \mathrm{Cov}((H - \langle H\rangle)^2, \delta H)\,. \tag{30}$$

Here the covariance is understood in the first-order spectral perturbation sense, i.e. through the induced energy shifts $\delta E_n = \langle n \mid \delta H \mid n\rangle$. The leading covariance term in Eq. (29) can be bounded by Cauchy–Schwarz. Since $\sigma(H) \;=\; \frac{\sqrt{C}}{\beta}$, one obtains $\beta^2 |\mathrm{Cov}(H, \delta H)| \;\lesssim \beta\sqrt{C}\, \|\delta H\|$. Similarly, the mixed central moment satisfies the estimate Eq. (31),

$$\left|\kappa_{2,1}\right| \;\lesssim \mathrm{Var}(H)\, \|\delta H\| \;=\; \frac{C}{\beta^2}\, \|\delta H\|\;\;, \tag{31}$$

up to numerical factors of order unity. Inserting the backward error-scale $\|\delta H\| \sim u\, W$ gives the correlated heat-capacity error estimate

$$|\Delta C| \sim |\delta C| \;\lesssim\; \beta\, u\, W \left(\sqrt{C} + C\right). \tag{32}$$

For the gapped finite systems considered here, $C$ is exponentially suppressed at low temperature. Hence both $\beta\sqrt{C}$ and $\beta C$ vanish as $T \to 0$, and the explicit factor of $\beta$ in Eq. (32) does not imply a low-temperature precision loss. This is consistent with the observed $10^{-7}$–$10^{-8}$ FP32-vs-FP64 error level.

Taken together, the estimates for $\Delta\chi$ and $\Delta C$ provide a backward-error-based explanation for the observed FP32 error scale. In the systems tested here, the precision error is consistent with a scale set primarily by $u\,W$, up to implementation-dependent factors. This explains why the observed FP32-vs-FP64 deviations remain essentially independent of the Hilbert-space dimension and far below the stochastic FTLM uncertainty. The same reasoning applies more generally to thermodynamic quantities that can be written as weighted sums over Ritz energies, or equivalently as integrals over the cumulative Lanczos weight $F(E)$. A finite-precision perturbation mainly shifts the spectral steps and redistributes weight within already converged clusters. Thus, for observables that commute with $H$, the expected precision error remains governed by the same operator-level scale. Finally note that the estimates above should not be understood as strict error bounds: they rely on a linear-response expansion around the FP32-effective Hamiltonian and use $u\,W$ as a worst-case perturbation scale. The numerical results of section 3.1 show that the actual FP32 errors lie well below these envelopes.

**3.4 Performance.** The dominant computational cost of FTLM is the sparse matrix-vector product (SpMV) inside the Lanczos iteration, which must be performed $N_L$ times for every random vector in every magnetization sector. Typical choices are $N_L = 100 - 200$ Lanczos steps and $R = 50 - 100$ random vectors per sector. We compare five variants on identical hardware: a CPU batched Lanczos implementation with CLT (FP64, 24 OpenMP threads), and four GPU variants in FP32 covering CLT and CR access patterns in single-vector and batched form. Here batched means that $B$ independent Lanczos chains are processed together so that the row-local Hamiltonian work is shared. Table 3 reports end-to-end timings for the CPU reference and the two batched GPU variants, while single-vector timings are used only to assess the per-SpMV cost of the access patterns. The batch size $B$ is chosen adaptively to balance vector-memory traffic, reuse of lookup data in cache, and the 20 GB device-memory limit.

For intermediate system sizes such as the $s = 3/2$ icosahedron, the CLT array (≈ 4 MB) is small enough to fit into the 48 MB L2 cache of the RTX 4000 Ada and to benefit from cache reuse, producing measurably higher throughput than a full lookup table that would have to be streamed from VRAM. For the $s = 1/2$ icosidodecahedron the CLT grows to 268 MB and exceeds the L2 capacity. It is still 16× smaller than the corresponding full lookup table (4.3 GB) and remains a natural choice whenever the table fits comfortably in VRAM. The CR kernel avoids the lookup altogether at the cost of a per-SpMV digit decomposition, and we examine its trade-off against the CLT below.

Table 3: Scaling study for the Heisenberg icosahedron and icosidodecahedron. Wall-clock times are for the timed Lanczos loop with $R \times N_L$ SpMVs per sector, $R = 24$ for the icosahedron and $R = 8$ for the icosidodecahedron. The $s = 2$ icosahedron and the icosidodecahedron timings refer to the $M = 0$ sector only; the $s = 1$ and $s = 3/2$ icosahedron timings span the full sum over $M \geq 0$. CPU timings use a batched-Lanczos OpenMP kernel with CLT ($B = 8$, FP64, 24 cores). The speedup reports CPU timing divided by the faster of the two GPU kernels.[a]

| System | $\mathcal{D}_{M=0}$ | CPU (s) | GPU-CLT (s) | GPU-CR (s) | Speedup |
|---|---|---|---|---|---|
| Icosahedron, $s = 1$ | $7.4 \cdot 10^4$ | 1.2 | 0.36 | 0.21 | 5.9× |
| Icosahedron, $s = 3/2$ | $1.7 \cdot 10^6$ | 191 | 14.8 | 14.6 | 13.1× |
| Icosahedron, $s = 2$ $(M = 0)$ | $2.0 \cdot 10^7$ | 381 | 36.7 | 34.8 | 10.9× |
| Icosidodecahedron, $s = 1/2$ $(M = 0)$ | $1.55 \cdot 10^8$ | 1180 | 124 | 135 | 9.5× |

[a] Batch sizes are chosen based on the heuristics described in the main text. GPU-CLT: $B = 8$ for the $s = 1$ icosahedron, $B = 4$ for the other systems; for GPU-CR: $B = 8$ for the $s = 1$ and $s = 3/2$ icosahedron, $B = 4$ for the other two systems.

The main performance trend is clear. Once the sector dimension exceeds roughly $10^6$, the GPU bandwidth advantage becomes visible and the speedup settles near one order of magnitude. The largest speedups are 13.1x for the full $s = \frac{3}{2}$ icosahedron calculation, 10.9x for $M = 0$, and 9.5x for the $s = \frac{1}{2}$ icosidodecahedron $M = 0$ sector. A separate $M = 0$ run for the $s = \frac{3}{2}$ icosahedron gives a 12.2x speedup, confirming that the comparison between full-sector and single-sector calculations does not bias this conclusion. The absolute ratio should not be interpreted as architecture-independent: it reflects the memory bandwidth, cache hierarchy, FP32 data volume, and latency hiding of the tested workstation GPU. More importantly, the matrix-free GPU representation reduces storage from the multi-gigabyte compressed-sparse representation to a basis, lookup structure, and Lanczos workspace that fit comfortably within the available device memory.

Within the GPU group, the batched CLT and CR kernels give similar end-to-end timings. The two strategies trade memory for arithmetic: CLT is faster per lookup when the compressed table fits in cache or device memory, whereas CR avoids the lookup table and reinvests the saved memory into the Lanczos workspace. In single-vector runs, where digit-decomposition work cannot be amortized over a batch, CLT is clearly faster; in batched runs the difference largely disappears. The practical conclusion is therefore that CLT is the preferred default when its table fits comfortably in GPU memory, while CR is the natural fallback when the state-to-index map itself becomes the limiting auxiliary storage.

Table 4: Analytic GPU memory budget for the CLT-based batched Lanczos kernel in the $M = 0$ sector of each system. Columns scale as $(\mathcal{D}/32) \cdot 8$ bytes (CLT array), $\mathcal{D}_0 \cdot 4$ bytes (basis array), and $3B\mathcal{D}_0 \cdot 4$ bytes (Lanczos workspace, i.e. three FP32 work vectors per batched chain). For the $s = 2$ icosahedron, the CLT and basis-array entries match the CLT column of Table 2.

| System | CLT array (MB) | Basis array (MB) | Lanczos vectors (MB) | Total (MB) |
|---|---|---|---|---|
| Icosahedron, $s = 3/2$ | 4 | 6.8 | 82 @ $B = 4$ | ≈ 93 |
| Icosahedron, $s = 2$ | 61 | 78 | 940 @ $B = 4$ | ≈ 1080 |
| Icosidodecahedron, $s = 1/2$ | 268 | 620 | 7440 @ $B = 4$ | ≈ 8330 |

The icosidodecahedron is the most demanding benchmark considered here. With $R = 8$, the batched kernels complete in 124 s (CLT) and 135 s (CR), compared with 1180 s for the 24-core CPU reference. This case demonstrates that sectors of order $10^8$ basis states can be treated on a single 20 GB workstation GPU. As historical context, an FTLM calculation of the same $M = 0$ sector of the icosidodecahedron was reported in 2010 [43], using $R = 20$ and $N_L = 100$, requiring approximately one day on 510 Itanium II cores with `spinpack`. Scaled linearly in $R$, our 24-core CPU reference would complete the equivalent calculation in ~50 minutes (~5 minutes on the GPU). This comparison reflects hardware and compiler evolution over 16 years, rather than a change in the underlying matrix-free access pattern and should therefore be read only as an order-of-magnitude marker. The meaningful benchmark remains the GPU/CPU comparison performed here on the same hardware generation.

## 4. Summary and outlook

We have presented a GPU implementation of the finite-temperature Lanczos method for Heisenberg spin clusters. The Hamiltonian action is evaluated directly in a row-wise gather formulation, so that each GPU thread accumulates one output-vector element without atomic updates. This matrix-free design shifts the dominant memory requirement from Hamiltonian storage to the basis representation, Lanczos work vectors, and the state-to-index map.

Two state-to-index strategies have been developed or adapted to GPU execution in this work. The first is a compressed lookup table (CLT) that represents the map by a bit mask and block-wise prefix counts, reducing the memory footprint by a factor of 16 relative to a full lookup table while preserving a fixed and branch-light lookup sequence. The second is a GPU-adapted variant of a combinatorial ranking (CR) scheme [16], in which the original data-dependent inner scan is replaced by a single lookup into a precomputed cumulative dimension table that is staged in shared memory, eliminating the basis array and any per-label index structure. Both schemes make the state-to-index translation suitable for GPU execution and together allow sector dimensions relevant to FTLM applications to be treated within the memory of a workstation GPU. They occupy complementary points in a memory-throughput trade-off: the CLT delivers the highest per-lookup throughput as long as its compressed table fits in device memory, whereas the GPU-adapted CR dispenses with the lookup table altogether and is therefore the natural choice whenever auxiliary storage for the state-to-index map becomes the dominant memory limitation.

We have also assessed the numerical effect of using single precision for the Lanczos recursion on the GPU. Across the benchmark systems considered here, the FP32–FP64 differences in heat capacity and magnetic susceptibility remain several orders of magnitude below the stochastic uncertainty of the FTLM trace estimator at typical sample sizes. The additional ghost Ritz values observed in FP32 mainly redistribute spectral weight within already converged clusters and do not lead to visible changes in the observables. These results indicate that, for the thermodynamic quantities studied here, FP32 arithmetic is sufficient for the dominant part of the FTLM calculation when the small tridiagonal eigenproblems are solved in double precision; other quantities, especially dynamical correlation functions or observables that do not commute with the Hamiltonian, may require separate tests.

The performance benchmarks show that the GPU implementation provides substantial speedups over optimized multicore CPU calculations once the problem is large enough for memory traffic and lookup overhead to dominate. The largest examples demonstrate FTLM calculations for Hilbert-space sectors with dimensions of order $10^8$. The present implementation uses only total-magnetization symmetry and is restricted to a single GPU. Additional point-group or translational symmetries would reduce sector dimensions, whereas still larger sectors will require a distributed-memory or multi-GPU formulation in which the basis, work vectors, and possibly the lookup structure are partitioned across devices. The same matrix-free strategy can be extended to more general spin Hamiltonians, including bond-dependent couplings and anisotropic exchange terms [44], provided that the local action on product-basis states remains inexpensive. Further improvements may come from combining the present lookup strategy with additional symmetry adaptation, multi-GPU execution, or mixed-precision variants in which storage and arithmetic precision are chosen separately.

**Code availability.** The `MATLAB/CUDA` source code, including example input files and the scripts required to reproduce the benchmark tables, is openly available at https://github.com/ghasdeke/ftlm-gpu under the Apache-2.0 license. The specific version used in this work (v1.0.0) is permanently archived on Zenodo with DOI 10.5281/zenodo.20378647 (https://doi.org/10.5281/zenodo.20378647).

**Acknowledgments.** S.G.T. was supported by the Deutsche Forschungsgemeinschaft (DFG) under Project 535298924.